\documentclass[aps,prl,twocolumn,footinbib,showpacs]{revtex4}
\usepackage{amssymb}
\usepackage{amsmath}    
\usepackage{mathrsfs}  
\usepackage{amsfonts}
\usepackage{graphicx}   
\usepackage{epstopdf}
\usepackage{bm}        
\usepackage{color}

\def\bs{\begin{subequations}}
\def\es{\end{subequations}}
\def\aa{\begin{align}}
\def\ab{\end{align}}
\def\ba{\begin{eqnarray}}
\def\ea{\end{eqnarray}}
\def\be{\begin{equation}}
\def\ee{\end{equation}}
\newcommand{\la}{\langle}
\newcommand{\ra}{\rangle}

\begin{document}

\title{Statistical Mechanics of Multistability in Microscopic Shells}

\author{Ee Hou Yong$^1$ and L. Mahadevan$^{1,2}$}
\affiliation{$^1$ Department of Physics, Harvard University, Cambridge, Massachusetts 02138\\
$^2$ School of Engineering and Applied Sciences, Kavli Institute for Nano-bio Science and Technology, Harvard University, Cambridge, MA 02138}
\date{\today}

\begin{abstract}
Unlike macroscopic multistable mechanical systems such as snap bracelets or elastic shells that must be physically manipulated into various conformations, microscopic systems can undergo spontaneous conformation switching between multistable states due to thermal fluctuations,  Here we investigate the statistical mechanics of shape transitions in small elastic elliptical plates and shells driven by noise. By assuming that the effects of edges are small, which we justify exactly for plates and shells with a lenticular section, we decompose the shapes into a few geometric modes whose dynamics are easy to follow. We use Monte Carlo simulations to characterize the shape transitions between conformation minima as a function of noise strength, and corroborate our results using a {\it Fokker-Planck} formalism to study the stationary distribution and the mean first passage time problem. Our results are applicable to objects such as such as graphene flakes or protein $\beta$-sheets, where fluctuations, geometry and finite size effects are important.
\end{abstract}


\maketitle


Many macroscopic elastic systems in nature and engineering are multistable, i.e. they can transition from one stable state to another in response to external perturbations; examples range from hair clips and slap bracelets \cite{Seffen2011, Giomi2011} to exotic morphing micro air-vehicles \cite{Abdulrahim2005}. Elastic multistability is also seen in microscopic systems, such as protein $\beta$-sheets which undergo conformational changes from a closed to an open state \cite{Oster2003, Branden1999}, graphene flakes \cite{Morgen2010}, composite lipid bilayers \cite{Zimm2003} etc. where geometry and finite size effects are relevant and transitions between metastable states are effected by thermal fluctuations. Although molecular dynamics can provide an accurate description of these transient behaviors, the computational cost makes these approaches unfeasible except for the smallest of systems. There is thus a need for {\it coarse-grained} dynamical models for the statistical mechanics of small microscopic plates and shells that focuses on the kinetics of and transitions between different conformations in terms of models for the coupled nonlinear {\it Langevin} dynamics of the modes that captures the qualitative trends. In this letter, we focus on such a minimal model that accounts for the role of fluctuations, finite size effects and geometry in spatially inhomogeneous multistable elastic systems.

\begin{figure}[htbp]
\centering
\includegraphics[width=3.4in]{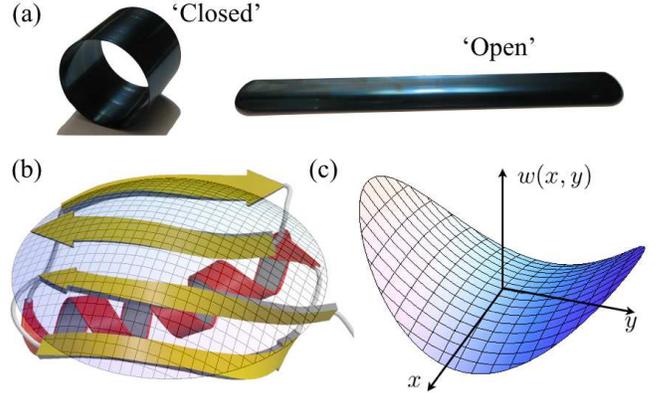}
\caption{(Color online)  (a) `Closed' and `open' state of a slap bracelet. (b) Model the shape of a microscopic sheet such as $\beta$-sheet by a differentiable coarse-grained map $w(x,y)$ (see Eq.~(\ref{eq:w})) with two bending modes $\kappa_x$, $\kappa_y$ and a twisting mode $\kappa_{xy}$. (c) Saddle with $\vec{\kappa} = (\kappa_x, \kappa_y, \kappa_{xy})^T = (-1,1,0)^T$. }
\label{fig:beta}
\end{figure}

Thin elastic plates and shells, with a thickness that is much smaller than their span, can be characterized in terms of the in-plane and out-of-plane deflection modes. While the general theory for such a low-dimensional object is mathematically formidable \cite{Landau}, the large deflection over-damped dynamical behavior of thin inhomogeneous shallow shells may be described by a generalization of the  F\"oppl-von K\'arm\'an theory, sometimes known as the Donnell-Mushtari-Vlasov equations.  For microscopic and nanoscopic plates and shells where inertial effects are dominated by viscous forces and thermal effects are important, these  read \cite{Mansfield1989}:
\ba
\Delta (B \Delta \tilde{w} ) + \eta h \partial_t w =  (1- \nu) [B, \tilde{w}] + [\Phi, w] + h\, \sigma(t, \vec{x}), \nonumber\\
\Delta (\Delta \Phi/h) = (1+\nu) \left[1/h, \Phi\right] +  \frac{E}{2} ([w^0,w^0] - [w, w] ) \quad
\label{eq:M} 
\ea
where $\Delta f = \partial_{xx}f + \partial_{yy}f$, $[f, g]\equiv \partial_{xx}f \,\partial_{yy}g + \partial_{yy}f\,\partial_{xx}g - 2 \,\partial_{xy}f\,  \partial_{xy}g $, $h(x,y)$ is the spatially inhomogeneous plate thickness, $\eta$ is the viscosity of the solvent, $\nu$ is the Poisson's ratio, $E$ is the Young's modulus, $B = E h^3/12(1 - \nu^2)$ is the bending stiffness of the plate made of a linear isotropic material, $w^0(x,y)$ is the spontaneous (or natural) out-of-plane displacement, $w(x,y)$ is the current out-of-plane displacement, $\tilde{w} = w-w^0$, $\Phi$ is the Airy stress potential whose second derivatives yield the components of the in-plane stress tensor, and $\sigma(t, \vec{x})$ is a random Gaussian forcing that we define later.  The first line in Eq.~(\ref{eq:M}) describes the {\it slow} dynamics associated with the balance of out-of-plane forces, while the second is a geometric compatibility relation that follows from the fast equilibration of in-plane  forces.  For later use, we point out that the elastic strain energy of the shell which characterizes its energy landscape is the sum of the bending and stretching of the middle surface \cite{Mansfield1967}:
\ba
U &=&  \iint \frac{B}{2}\{(\Delta \tilde{w})^2 - (1-\nu)[\tilde{w},\tilde{w}]\} dA\nonumber \\
&&+\iint\frac{1}{2Eh} \{(\Delta \Phi)^2 - (1+\nu) [\Phi, \Phi] \} dA. 
\label{eq:U}
\ea
Rather than proceeding by decomposing the deformations of the shell in terms of Fourier (or some other) modes without worrying about the boundaries, as is traditional in infinite systems, for small plates and shells that fluctuate freely, we must account for the boundary conditions that state that the edges are free of torques and forces, so that \cite{Mansfield1989},
\ba
&&B[ \partial_{nn} + \nu (\partial_{ss} + \partial_s \psi \partial_n )] \tilde{w}|_\Omega = 0, \nonumber \\
&&B\{ \partial_n \Delta + (1-\nu)\partial_s[\partial_{ns} - \partial_s \psi \partial_s)]  \label{eq:BC} \\
&&\quad+ \partial_n B[ \partial_{nn} + \nu(\partial_{ss} + \partial_s \psi \partial_n)] \nonumber \\
 &&\quad+ 2(1-\nu)\partial_s B( \partial_{ns} - \partial_s \psi \partial_s)\} \tilde{w}|_\Omega = 0,\nonumber 
\ea
where $n$ and $s$ denote the normal and tangential direction along the plate/shell boundary  curve $\psi(s) \in \Omega$ with curvature $\partial_s \psi$. In general, the conditions (\ref{eq:BC}) lead to the existence of elastic boundary layers near the edge \cite{Mansfield1989} and thence make the solution of the governing equations difficult. However for a special kind of plate, the $L$-plate \cite{Mansfield1965, Mansfield1967}, with a {\it lenticular} section that vanishes along the boundary according to 
\be
h(x,y) = h_0 ( 1 - x^2/a^2 - y^2/b^2),
\ee
where $h_0$ is the thickness at the center and $a, b$ are the semi-major and semi-minor axes of the plate respectively, the   flexural rigidity $B = B_0 (1 - x^2/a^2 - y^2/b^2 )^3$ where $B_0 = Eh_0^3/12(1-\nu^2)$ decreases smoothly to zero at the boundaries, i.e. $B|_\Omega =  \partial_n B |_\Omega=  \partial_s B |_\Omega = 0$,
and thus (\ref{eq:BC}) automatically vanish, eliminating the boundary layers which would otherwise occur \cite{Mansfield1989}.

This simplification can be used together with a geometric perspective to describe inhomogeneous deformations of the $L$-plate/shell as follows: given a point $p$ of a surface $S$, we can parametrize the local neighborhood with a differentiable map $w(x,y)$ with $w(0,0) = \partial_x w(0,0) = \partial_y w(0,0) = 0$ \cite{DoCarmo1976}: 
\be
w(x, y) = c [ \kappa_x (x^2 -a^2/6) + 2\kappa_{xy} xy + \kappa_y (y^2 -b^2/6 )],
\label{eq:w}
\ee
where $c = -\frac{h_0}{2ab\sqrt{1-\nu^2}} \sqrt{ 4+2\nu + 5 (\zeta^2 + \zeta^{-2})}$ and $\zeta = b/a$ is the aspect ratio of the plate/shell, and the constant terms are included to ensure $\int \int_A h w dx dy = 0$. The nondimensional curvatures $\kappa_x, \kappa_y$ correspond to bending in the $x$ and $y$ directions and $ \kappa_{xy}$ corresponds to the twisting mode. For small plates, these 3 elementary modes are sufficient (in a coarse-grained sense) to characterize the shape completely. We choose the hitherto unprescribed form of the random Gaussian forcing to be of the form:
\be
\sigma(t, \vec{x}) = c_1 [ \gamma_x (x^2 -a^2/6) + 2\gamma_{xy} xy + \gamma_y (y^2 -b^2/6 )],
\ee
 to allow the white noise to also be spatio-temporally decoupled, with $c_1 = 12B_0 c/h_0 a^2 b^2$, and $\gamma_{i}$'s ($i = x, y, xy$) being independent stochastic noise terms that are Gaussian distributed with zero mean $\la \gamma_{i} (t) \ra = 0$, and covariance $\la\gamma_{i} (t)\gamma_{j} (t') \ra = 2\alpha {D}_{ij} \delta(t - t')$ with ${\bm D}$ being the diffusion tensor that characterizes the strength of the thermal noise and $\alpha = \eta h_0a^2b^2/12B_0$. 

Assuming that the Airy stress function has the form $\Phi = B \beta(t)$ for an elastic shell with spontaneous curvatures described in terms of the constants $\kappa_x^0, \kappa_{xy}^0, \kappa_y^0$, and substituting this form along with the deflection $w(x,y)$ in Eq.~(\ref{eq:w}) into the governing equations (\ref{eq:M}) yields the resulting Langevin equation for the evolution of the curvature components $\vec{\kappa}  = (\kappa_x, \kappa_y, \kappa_{xy})^T$:
\be
\dot{\vec{\kappa}} = \vec{v}(\tau, \vec{\kappa} ^0, \vec{\kappa}(\tau)) + \vec{\gamma}(\tau),
\label{eq:langevin}
\ee
where  $\vec{v}(\tau, \vec{\kappa} ^0, \vec{\kappa}(\tau)) = {\bm A}( \vec{\kappa} ^0- \vec{\kappa}(\tau) )+ \beta \vec{f}(\tau)$ is the curvature velocity, $\dot{f} \equiv \partial f/\partial\tau$, $\tau =\alpha^{-1} t$, 
\ba
\beta(\tau) = (\kappa_{x}^0 \kappa_{y}^0 -  (\kappa_{xy}^0)^2) - (\kappa_{x}(\tau) \kappa_{y}(\tau) -  \kappa_{xy}^2(\tau)), \\
 {\bm A} = \left(\begin{array}{ccc}5\zeta^2 + \nu  &1 + 5 \nu \zeta^2 & 0 \\  1 + 5 \nu \zeta^{-2} &  5\zeta^{-2} + \nu & 0 \\0 & 0 & 4(1- \nu)\end{array}\right), \\
\vec{f}(\tau) = \left(\begin{array}{c} \kappa_x + 5 \zeta^2 \kappa_y \\ 5\zeta^{-2}\kappa_x + \kappa_y \\- 4 \kappa_{xy}\end{array}\right), \, \vec{\gamma}(\tau)= \left(\begin{array}{c}\gamma_x \\\gamma_y \\\gamma_{xy}\end{array}\right).
\ea

We note that the curvature velocity $\vec{v} = -{\bm \mu} \nabla_\kappa U$, where $\nabla_\kappa \equiv \partial/\partial \vec{\kappa}$, and the elastic energy of the $L$-shell, given by Eq.~(\ref{eq:U}) reads
\ba
&&U= U_0[ -2(1-\nu)\{(\kappa_x - \kappa_x^0)(\kappa_y - \kappa_y^0) - (\kappa_{xy} - \kappa_{xy}^0)^2\}  \nonumber \\
&&\quad +  (\kappa_x+\kappa_y - \kappa_x^0 - \kappa_y^0)^2 + \beta^2],
\label{eq:U1}
\ea
with $U_0 = \pi B_0 a b c^2/2$. Then, it follows that 
\be
{\bm \mu} =  \left(\begin{array}{ccc}5\zeta^2/2U_0  & 1/2U_0 & 0 \\  1/2U_0 &  5\zeta^{-2}/2U_0 & 0 \\0 & 0 & 1/U_0\end{array}\right),
\ee
so that the diffusion tensor  ${\bm D} = {\bm \mu} k_B T$, following from the Einstein-Smoluchowski relation. This allows us to visualize the geometrical dynamics of $\vec{\kappa}$ evolving on a landscape determined by the elastic energy (\ref{eq:U1}). 

When the spontaneous curvature $\vec{\kappa}^0$ vanishes, we find that the elastic energy adopts a simplified form
\be
U = U_0[(2H)^2 - 2(1-\nu) K + K^2],
\ee
with the mean curvature $H = (\kappa_x+\kappa_y)/2$  and the Gaussian curvature $K = \kappa_x\kappa_y - \kappa_{xy}^2$  having a single minimum at $H = K = 0$, i.e. a planar state. Since the governing curvature potential given by Eq.~(\ref{eq:U1}) is quartic in $\kappa$ ($\beta$ is quadratic in $\kappa$), there can be a maximum of three extrema. In the naturally flat plate case, the remaining two roots are complex. Thus, the plate follows a simple stochastic motion around the single stable flat state. Likewise, a naturally parabolic plate ($K_0  = 0$, $H_0 \neq 0$) has only one minimum and two complex roots, and the stochastic dynamics is similar to that for a flat plate. When the plate is naturally hyperbolic ($K_0 < 0$), the system possesses one minimum and two saddles and hence it is {\it monostable}. However, when the plate is elliptic ($K_0 > 0$), the system is {\it bistable}, with two minima, $\vec{\kappa}^0$ (global), $\vec{\kappa}^m$ (local) separated by a saddle $\vec{\kappa}^{s}$. If the initial configuration of the $L$-shell is near $\vec{\kappa}^m$, the system will be trapped in this local basin of attraction unless the temperature is high enough, in which case it will sample the global minimum $\vec{\kappa}^0$ as well. Since only the naturally elliptic plate is {\it bistable}, we will focus solely on this case for the rest of the paper. 

\begin{figure}[htbp]
\centering
\includegraphics[width=3.4in]{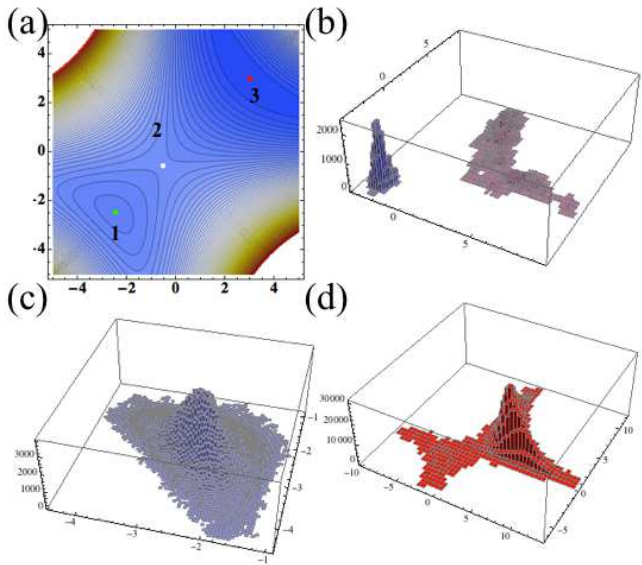}
\caption{(Color online)  (a) Contour plot of the elastic energy as function of $\kappa_x$ and $\kappa_y$. Point 1 is the local minimum, point 2 is the saddle point and point 3 is the global minimum. (b) Two separate stochastic realizations with plate initially at point 1 and $T=0.1/1$ (blue/red histogram). (c) Histogram of equilibrium statistics consisting of $10^3$ stochastic realizations under cold conditions ($T=0.1$) with the plate initially at point 1. (d) Same as (c) but under hot conditions ($T=1$).  (See movie.)  }
\label{fig:circular}
\end{figure}

To understand the stochastic dynamics of conformational switching in naturally elliptic $L$-plates and shells, we perform Monte Carlo simulations by discretizing Eq.~(\ref{eq:langevin}) in the time domain, with $\tau = N\Delta \tau$, and write:
\be
\vec{\kappa}_{n+1} = \vec{\kappa}_{n} + \vec{v}( \vec{\kappa}^0, \vec{\kappa}_{n}) \Delta \tau + \sqrt{2\Delta \tau}{\bm L} \,\vec{\xi}_n, \, n \in \mathbb{Z},
\label{eq:update}
\ee
where $\vec{v}( \vec{\kappa}^0, \vec{\kappa}_{n})$ is the {\it deterministic} velocity term evaluated at $\vec{\kappa}_{n}$, ${\bm L}$ is the Cholesky decomposition of the diffusion tensor ${\bm D}$ and $\vec{\xi}_n$ is a random vector with entries drawn from standard normal distribution \cite{foot1}.  We assume that the $L$-plate is initially at the local minimum configuration $\vec{\kappa}^m$ and let the curvatures evolve in time via Eq.~(\ref{eq:update}). Each stochastic realization consist of $10^4$ Monte Carlo time steps, of which the first 80\% is devoted to equilibration and there is a total of $10^3$ realizations for each temperature considered. This allows us to understand (1) the equilibrium statistics of the $L$-plate and (2) the mean survival time of the metastable initial state $\vec{\kappa}^m$, namely, the time it takes the plate to go from the metastable initial state $\vec{\kappa}^m$ to the stable state $\vec{\kappa}^0$. 

For the case of a shell with a circular boundary, i.e. $\zeta = 1$, and natural curvature $\vec{\kappa}^0 = (3,3,0)^T$ corresponding to a naturally elliptical shell with $K_0=3, H_0=9$, neglecting the dynamics of $\kappa_{xy}$ for the moment, i.e. with principal directions are invariant in time, we can view the stochastic dynamics of the two principal curvatures $\kappa_x$ and $\kappa_y$. The system contains two minima and a saddle point as shown in Fig.~\ref{fig:circular}(a). If the shell curvature is initially at the local minimum $\vec{\kappa}^m$, when the scaled temperature $T = 0.1$, it remains distributed close to this potential minimum. However, when $T = 1$, the shape will diffuse into the basin of the global minimum and the distribution of the curvature is very spread out as shown in Fig.~\ref{fig:circular}(b). At equilibrium (data from $10^3$ runs), the shape remains distributed close to its natural minima when it is cold, but becomes spread out when it is hot as shown in Figs.~\ref{fig:circular}(c), (d) respectively.

\begin{figure}[!h]
\centering
\includegraphics[width=3.4in]{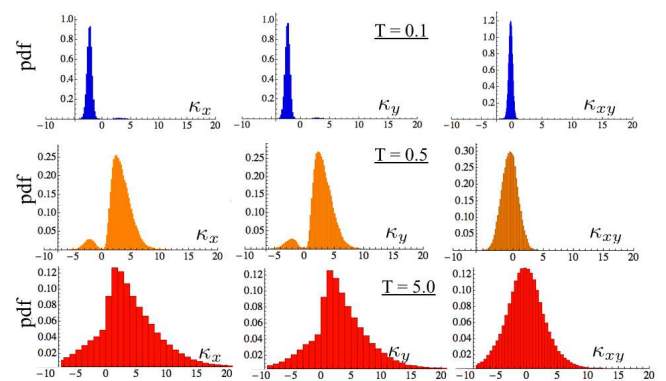}
\caption{(Color online) (a) The steady-state pdf $p(\kappa_x$), $p(\kappa_y$) and $p(\kappa_{xy}$) respectively at cold conditions ($T= 0.1$) (b) mild conditions ($T= 0.5$) and (c) hot conditions ($T= 5$) for an circular plate with $\vec{\kappa}^0 = (3,3,-1)^T$ and $\vec{\kappa}^m = (-2.2, -2.2, -0.2)^T$. }
\label{fig:prob}
\end{figure}

Next, we consider a shell with a circular boundary, i.e. $\zeta = 1$, with $\vec{\kappa}^0 = (3,3,-1)^T$ and initial curvature $\vec{\kappa}^m = (-2.2,-2.2,-0.2)^T$. Since $K_0 = 3$ and $H_0 = 8$, this is an naturally elliptic plate. The energy landscape contains two minima denoted by $\vec{\kappa}^0$ and $\vec{\kappa}^m$ and a saddle point. In Fig.~\ref{fig:prob}, we plotted the steady state probability density function (pdf) for $\kappa_x, \kappa_y$ and $\kappa_{xy}$ (conditional on the plate being initially at $\vec{\kappa}^m$) from $10^3$ stochastic realizations. Under cold conditions ($T = 0.1$), the equilibrium curvatures of the plate remains distributed close to its local minimum $\vec{\kappa}^m$ as shown in Fig.~\ref{fig:prob}(a). In this case, the system is trapped in the local minimum and thermal effect is negligible. As we increase the thermal noise ($T = 0.5$), the distribution of the steady state plate curvatures spreads and spend a significant amount of time in both minima as reflected by the bimodal pdf in Fig.~\ref{fig:prob}(b). In this case, the thermal fluctuations cause the system to diffuse out of the local minimum into the global minimum. At even hotter conditions ($T = 5$), the equilibrium conformations of the plate have a large spread centered around its global minimum $\vec{\kappa}^0$ as shown in Fig.~\ref{fig:prob}(c). In this case, the thermal noise causes the system to become very spread out, similar to a pure diffusive process. 


In order to understand these results qualitatively, we switch from a stochastic Langevin formalism to a deterministic Fokker-Planck formalism, writing the probability $p(\vec{\kappa}, \tau)$ of finding the plate with curvature $\vec{\kappa}$ as given by the continuity equation, $\dot{p} + \nabla_\kappa \cdot\vec{j} = 0$, with probability current $\vec{j} = \vec{v}\, p - {\bm D} \nabla_\kappa p$, leading to the Fokker-Planck equation \cite{Coffey1996}, 
\be
\dot{p}(\vec{\kappa}, \tau) = - \nabla_\kappa \cdot ( \vec{v} \, p(\vec{\kappa}, \tau) - {\bm D} \nabla_\kappa p(\vec{\kappa}, \tau)),
\ee
with initial condition $p(\vec{\kappa}, 0) = \delta(\vec{\kappa}-\vec{\kappa}^m)$. For stationary processes, $\vec{j} = 0$, and the steady state probability density follows a Boltzmann distribution, $p_{\text{eq}} \propto \exp(-U/k_BT)$. The shell, in switching between the two minima, spends most of its time in the neighborhood of the saddle point, with the escape rate through the saddle depending on the Hessian  of the elastic energy around it \cite{Langer1969}, given by: 
\begin{widetext}
\be
{\bm H} = \left(\frac{\partial^2 U}{\partial\kappa_i\partial\kappa_j}\right) = 2U_0
 \left(\begin{array}{ccc} 1+ \kappa_y^2 & \nu - \beta + \kappa_x\kappa_y & -2 \kappa_y\kappa_{xy} \\ \nu-\beta+\kappa_x\kappa_y & 1+ \kappa_x^2 & -2\kappa_x\kappa_{xy} \\ -2 \kappa_y\kappa_{xy}& -2\kappa_x\kappa_{xy} & 2(1-\nu + \beta) + 4\kappa_{xy}^2\end{array}\right).
\ee
\end{widetext}
Denoting the Hessian at the saddle by ${\bm H}^s$ and its counterpart at the local minimum by ${\bm H}^m$ respectively, we write the transition matrix ${\bm M} = -{\bm \mu} {\bm H}^s$ \cite{Coffey1996,Langer1969,MaSt1993} which has a single positive eigenvalue $\lambda_+$ that characterizes the barrier crossing rate. In the over-damped limit, the inverse escape rate $\Gamma^{-1}= \Pi$, the mean first passage time, is given by \cite{Coffey1996, Langer1969, MaSt1993}
\be
\Pi \approx \frac{2\pi}{\lambda_+} \sqrt{\frac{|\det[ {\bm H}^s]|}{\det[ {\bm H}^m]} } \exp(\Delta U/k_BT),
\label{eq:MFPT}
\ee 
where $\Delta U = U^s - U^m$ is the energy difference between the saddle and local minimum. The exponential factor represents the probability that the energy will exceed that of the barrier when the system is in thermal equilibrium and is the only term in Eq.~(\ref{eq:MFPT}) that is temperature dependent. 

\begin{figure}[htbp]
\centering
\includegraphics[width=3in]{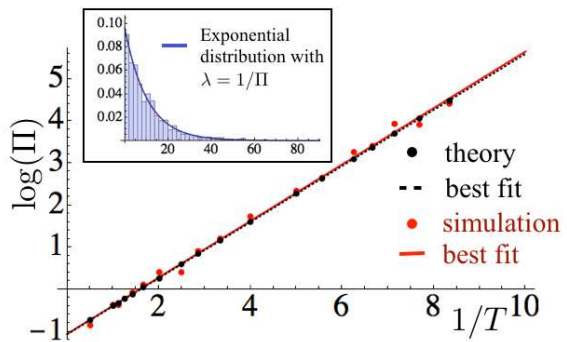}
\caption{(Color online) Plot of $\log(\Pi)$ as a function of $1/T$. $\Pi$ is the MFPT of the microscopic plate to escape over a barrier from initial curvature $\vec{\kappa}^m$ to the global minimum $\vec{\kappa}^0$. The curves are lines of best fit to data/theory. Inset: Histogram of the FPT for $T = 0.2$. The FPT has an exponential distribution with parameter $\lambda = 1/\Pi \approx 0.1$ as shown in the blue curve. }
\label{fig:MFPT}
\end{figure}
To verify this asymptotic formula, for each temperature $T$, we perform $10^3$ Monte Carlo simulations of the FPT and then calculate the mean and standard deviation. In Fig.~\ref{fig:MFPT} we show that our simulations and our theoretical estimate in Eq.~(\ref{eq:MFPT}) agree well. Our simulations also allow us to determine the mean and standard deviation of the FPT  over the full range of temperatures investigated, which are best expressed as  $\Pi$ = E[FPT] $\approx \sqrt{\text{Var}[\text{FPT}]}$, with the distribution of FPT following an exponential distribution with parameter $\lambda = 1/\Pi$ \cite{foot2}.


Our study exposes the stochastic switching dynamics between conformations of mesoscale membrane-like objects whose deformations can be described in terms of just a small number of modes. By using an analytically tractable elastic model for weakly curved plates and shells, the $L$-plate/shell with a lenticular section, we avoid the complications of elastic boundary layers, and can reduce the spatiotemporal evolution of morphologies to the interplay between geometry and thermal fluctuations on a curvature energy landscape. This allows us to characterize the equilibrium statistics and mean survival time of the metastable states for elliptic shells using a combination of numerical simulation and simple analytical estimates.  Our model shows how geometrical shape transitions can be actuated by thermal noise in microscopic systems, and is amenable to experimental tests in a variety of physical and biological systems. 

We thank the Harvard-MRSEC DMR0820484, the Harvard-Kavli Nan-Bio Science and Technology Institute, and the MacArthur Foundation (LM) for support.

\appendix

\section{Appendix: Autocorrelation}

The autocorrelation \cite{Jenkins2008} of a random process describes the correlation between values of the process at different times, as a function of the two times or of the time difference and is useful for detecting non-randomness in data. It is a useful mathematical tool for finding repeating patterns, such as the presence of a periodic signal which has been buried under noise, or identifying the missing fundamental frequency in a signal implied by its harmonic frequencies. Given measurements, $X_1, X_2,\dots X_N$ at time $t_1, t_2, \dots, t_N$, the lag $k$ autocorrelation function is defined as
\be
R(k) = \frac{\sum_{i=1}^{N-k} (X_i - \bar{X})(X_{i+k} - \bar{X})}{\sum_{i=1}^{N} (X_i - \bar{X})^2}
\label{eq:CF}
\ee
\begin{figure*}[htbp]
\centering
\includegraphics[width=6in]{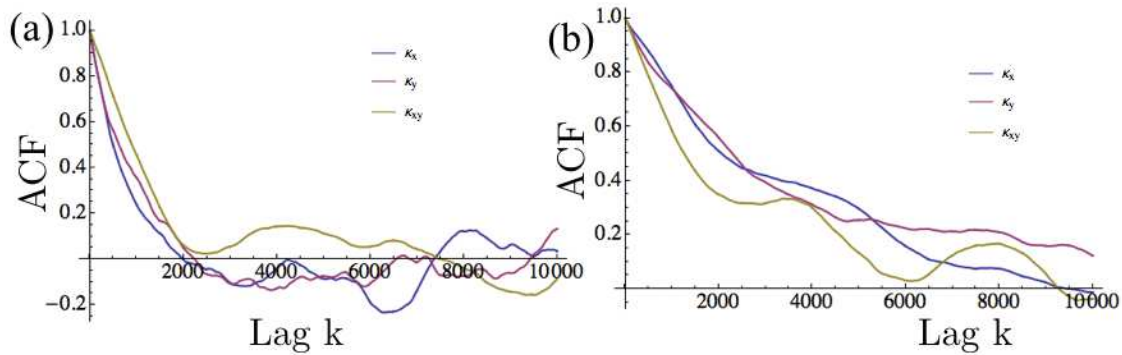}
\caption{ (a) Autocorrelation Function (ACF) as a function of lag $k$ for stochastic plate under cold conditions ($T=0.1$). Horizontal axis: Time lag $k = 1, 2, 3, \dots$ (b) Same as (a) but at $T=5$. }
\label{fig:ACF}
\end{figure*}

For our system, described by
\be
\vec{\kappa}_{n+1} = \vec{\kappa}_{n} + \vec{v}( \vec{\kappa}^0, \vec{\kappa}_{n}) \Delta \tau + \sqrt{2\Delta \tau}{\bm L} \,\vec{\xi}_n, \, n \in \mathbb{Z},
\ee
where $\vec{\kappa} = (\kappa_x, \kappa_y, \kappa_{xy})^T$, the ACF plot for the different $\kappa$'s starts with autocorrelation of 1 at lag 1 that gradually decreases with lag $k$. The decreasing autocorrelation is smooth with small oscillatory behavior and we say that such pattern exhibits a plot signature of ``high autocorrelation". Although $\vec{\kappa}_n$ depends only on $\vec{\kappa}_{n-1}$, the correlations at large lags are, nevertheless, nonzero. This should not be surprising because $\vec{\kappa}_{n-1}$ depends on $\vec{\kappa}_{n-2}$, and in turn $\vec{\kappa}_{n-2}$ on $\vec{\kappa}_{n-3}$, and so on, leading to an indirect dependence of $\vec{\kappa}_n$ on $\vec{\kappa}_{n-k}$ where $n > k$. 

\begin{figure*}[htbp]
\centering
\includegraphics[width=6in]{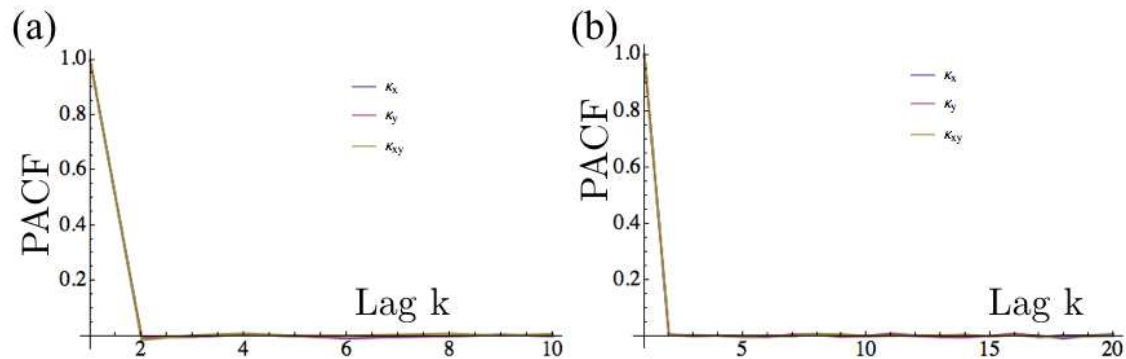}
\caption{ (a) Partial autocorrelation function (PACF) vs lag $k$ at $T=0.1$. (b) Same as (a) but at $T=5$.}
\label{fig:PACF}
\end{figure*}

Consider the conditional expectation $E(X_nX_{n-2}|X_{n-1})$ of an autoregressive process of order 1, i.e. AR(1) \cite{Jenkins2008} defined as, 
\be
X_n =\phi_1 X_{n-1} +Z_t,
\ee
where $X_n$ is a random process at time step $n$, $\phi_1$ is a constant and $Z_t$ is a white noise. That is, given $X_{n-1}$, what is the correlation between $X_n$ and $X_{n-2}$? It is clearly zero since $X_n$ is not influenced by $X_{n-2}$ given $X_{n-1}$. The partial autocorrelation between $X_n$ and $X_{n-k}$ is defined as the correlation between the two random variables with all variables in the intervening time $t-1, t-2, ... , t-k+1$ assumed to be fixed. Let $L_{X_1, X_2,\dots X_N}(X_{N+1})$ denote the best (in terms of minimizing MSE) linear predictor of $X_{N+1}$ based on $X_1, X_2,\dots X_N$.
\ba
\hat{X}_{N+1} &&= L_{X_1, X_2,\dots X_N}(X_{N+1}) \nonumber \\
&&= a_1 X_1 + \cdots + a_N X_{N},
\ea
where $a_1, a_2, \dots, a_N$ are determined by minimizing 
\be
E[({X}_{N+1}-\hat{X}_{N+1})^2].
\ee
The partial autocorrelation function (PACF) at lag $k$ of a stationary process$\{X_t\}$ is defined as \cite{Jenkins2008}
\be
\rho(1) = Corr(X_t, X_{t+1})
\ee
and
\begin{widetext}
\be
\rho(h) = Corr(X_t - L_{X_{t+1},..., X_{t+h-1}}(X_t),X_{t+h} - L_{X_{t+1},..., X_{t+h-1}}(X_{t+h}) 
\ee
\end{widetext}
for $h \ge 2$. The PACF at lag h may be interpreted as the correlation between $X_t$ and $X_{t+h}$ with the effect of the intermediate variables $X_{t+1},...,X_{t+h-1}$ ``filtered out". Clearly, for an AR($p$) process the partial correlation so defined is zero at lags greater than the AR order $p$. This fact is often used in attempts to identify the order of an AR process. For our simulation, we find that the PACF has a sharp cutoff at lag $k = 2$ and this is not surprising since our Langevin equation is a first order stochastic differential equation and hence an AR(1) process.


\begin{thebibliography}{}

\bibitem{Seffen2011}
K. A. Seffen and S. D. Guest, J. Appl. Mech. {\bf 78}, 011002 (2011).

\bibitem{Giomi2011}
L. Giomi and L. Mahadevan, Proc. R. Soc. A {\bf 468}, 511 (2011).

\bibitem{Abdulrahim2005}
M. Abdulrahim, H. Garcia, and R. Lind, J. of Aircraft {\bf 42}, 131 (2005).

\bibitem{Oster2003}
[4] S. Sun, D. Chandler, A. R. Dinner, and G. Oster, Eur Biophys J. {\bf 32}, 676 (2003).

\bibitem{Branden1999}
C. Branden and J. Tooze, {\it Introduction to Protein Structure} (Garland Publishing Inc., New York, 1999), 2nd ed.

\bibitem{Morgen2010}
T. Mashoff, M. Pratzer, V. Geringer, T. J. Echtermeyer, M. C. Lemme, M. Liebmann, and M. Morgenstern, Nano Lett. {\bf 10}, 461 (2010).

\bibitem{Zimm2003}
V. A. Frolov, V. A. Lizunov, A. Y. Dunina-Barkovskaya, A. V. Samsonov, and J. Zimmerberg, PNAS {\bf 100}, 8698 (2003).

\bibitem{Landau}
L. D. Landau, L. P. Pitaevskii, E. M. Lifshitz, and A. M. Kosevich, {\it Theory of elasticity} (Pergamon, NY, 1986), 3rd ed.

\bibitem{Mansfield1989}
E. H. Mansfield, {\it The bending and stretching of plates} (Cambridge University Press, Cambridge, UK, 1989), 2nd ed.

\bibitem{Mansfield1967}
G. Z. Harris and E. H. Mansfield, Phil. Trans. R. Soc. A {\bf 261}, 289 (1967).

\bibitem{Mansfield1965}
E. H. Mansfield, Proc. of the Royal Soc. A {\bf 288}, 396 (1965).

\bibitem{DoCarmo1976}
M. P. D. Carmo, {\it Differential geometry of curves and surfaces} (Prentice-Hall Inc., New Jersey, 1976).

\bibitem{Coffey1996}
W. Coffey, Y. P. Kalmykov, and J. T. Waldron, {\it The Langevin Equation} (World Scientific, Singapore, 1996). 

\bibitem{Langer1969}
J. Langer, Annals of Physics {\bf 54}, 258 (1969).

\bibitem{MaSt1993}
R. S. Maier and D. L. Stein, Phys. Rev. E {\bf 48}, 931 (1993). 

\bibitem{foot1}
Some interesting properties of the Langevin equation given by Eq. (14) can be found in Appendix.

\bibitem{foot2}
For comparison, the distribution of the FPT in the classic GamblerÕs ruin problem follows an inverse Gaussian distribution.

\bibitem{Jenkins2008}
G. E. P. Box, G. M. Jenkins, and G. C. Reinsel, {\it Time Series Analysis: Forecasting and Control}, 4th ed. (Wiley, New Jersey, 2008).

\end{thebibliography}

\end{document}